\documentclass{article}
\usepackage{spconf,amsmath,graphicx,hyperref}
\usepackage{spconf,amsmath,amssymb,graphicx}
\usepackage{booktabs}
\usepackage{cite}
\usepackage{microtype}
\usepackage{url}
\usepackage{xurl}
\usepackage{setspace}
\newcommand{\ours}{StateVC}

\title{Shared-State Local Translations for Training-Free Voice Conversion}
\name{
Yangyang Qu \qquad
Michele Panariello \qquad
Massimiliano Todisco \qquad
Nicholas Evans 
}

\address{EURECOM, Sophia Antipolis, France}
\begin{document}
\maketitle

\begin{abstract}
In one-shot training-free voice conversion (VC), the source and reference utterances may contain different linguistic content, so reliable frame-level correspondence between them cannot be assumed.
We propose \ours{}, which jointly defines a common set of local regions from pooled frame-level WavLM representations of the source and reference utterances; we refer to these regions as \emph{states}.
These shared states are obtained by fitting a pair-specific Gaussian mixture model to the pooled representations, without explicit source--reference frame matching.
Within each state, \ours{} estimates a source-to-reference mean shift in the original WavLM space.
Source-frame posterior probabilities then combine the state-specific shifts so that different frames can receive different local updates.
For the LibriSpeech one-shot protocol, \ours{} achieves the lowest word error rate (WER) and character error rate (CER) among the evaluated systems, at 8.01\% and 3.22\%, respectively, with a speaker similarity (SIM) of 0.9512.
It also achieves the highest mean perceived speaker similarity among the evaluated systems and the highest mean naturalness among the evaluated training-free systems.
\end{abstract}

\begin{keywords}
voice conversion, self-supervised speech representations, Gaussian mixture model, training-free conversion
\end{keywords}

\begin{figure*}[t]
    \centering
    \includegraphics[width=\textwidth]{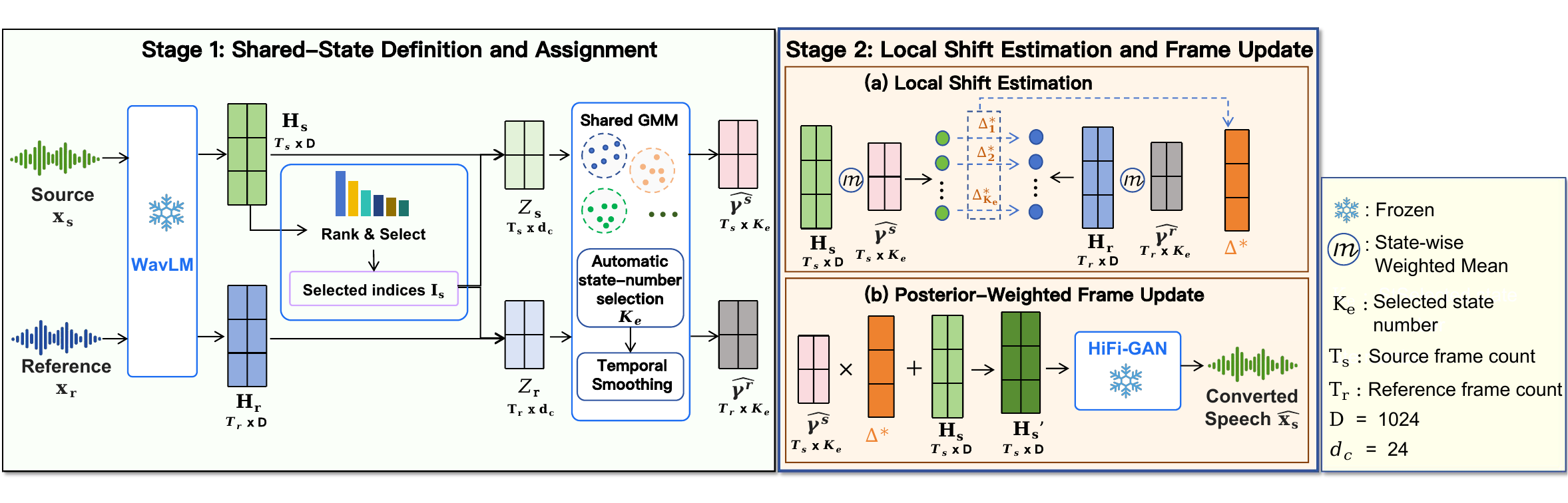}
    \caption{
    Overview of \ours{}.
    Source and reference representations jointly define a common set of shared states through a pair-specific GMM in a compact routing space.
    Source and reference means within each state define full-dimensional local shifts.
    Source-frame posterior probabilities combine these shifts before frozen HiFi-GAN synthesis.
    }
    \label{fig:overview}
\end{figure*}

\section{Introduction}
\label{sec:introduction}

Reference-based voice conversion (VC) aims to modify the speaker characteristics of a source utterance toward those of a reference utterance while preserving the source linguistic content~\cite{mohammadi2017overview}.
We consider one-shot VC with a single reference utterance, whose linguistic content may differ from that of the source; therefore, reliable frame-level correspondence is not assumed.
In training-free VC, conversion is performed by modifying pretrained speech representations without training a task-specific conversion network~\cite{baas2023knnvc,kamper2025linearvc,lobashev2025mklvc}.
We hypothesize that, for source frames with similar WavLM~\cite{chen2022wavlm} representations, the source-to-reference change can be approximated by a common shift.
This motivates defining local regions in representation space and estimating a separate source-to-reference shift within each region, rather than matching individual source and reference frames.

Existing VC methods based on self-supervised speech representations differ in how they establish source--reference relations and in whether they use a single mapping or multiple local mappings.
Nearest-neighbor and frame-based optimal transport methods explicitly relate source and reference frames~\cite{baas2023knnvc,asadulaev2024otvc,selitskiy2026dotvc}, while LinearVC fits a single linear transformation from nearest-neighbor-paired frames~\cite{kamper2025linearvc}.
MKL-VC estimates factorized distribution-level mappings~\cite{lobashev2025mklvc}, while USCF uses a shared low-rank representation with speaker-specific linear transformations~\cite{li2026uscf}.
Classical GMM-based VC learns mixture-dependent mappings from speaker-specific or parallel training data~\cite{stylianou1998continuous,toda2007voice}, whereas SSL-GMMVC obtains local affine mappings by first pairing source and reference self-supervised representations and then fitting a GMM~\cite{tanabu2026sslgmmvc}.
These approaches leave open whether local source-to-reference changes can be estimated without first pairing source and reference frames.

To address this problem, we introduce \ours{}, which pools source and reference representations and fits a single GMM for each utterance pair to define shared local regions.
We refer to these regions as \emph{states}.
This joint definition allows source and reference statistics to be estimated within the same state without first matching individual frames or separately estimated components.
For each state, \ours{} estimates a source-to-reference mean shift in the original WavLM space.
Source-frame posterior probabilities then combine these state-specific shifts, allowing the applied update to vary across frames.


Our contributions are:
\begin{itemize}

\item We propose \ours{}, a training-free one-shot VC method that jointly defines one set of local states shared by the source and reference, and uses state-specific mean shifts with posterior-weighted frame updates without explicit source--reference frame matching.

\item Controlled comparisons show that jointly defining one common set of states shared by source and reference yields lower WER and CER than defining them separately and matching them afterward.

\item Under the LibriSpeech one-shot protocol, \ours{} achieves the lowest WER/CER among the evaluated systems with a SIM of 0.9512, alongside the highest mean perceived speaker similarity and the highest mean naturalness among the evaluated training-free systems.
\end{itemize}

\section{Method}
\label{sec:method}

An illustration of \ours{} is provided in Figure~\ref{fig:overview}.
A frozen WavLM~\cite{chen2022wavlm} encoder extracts frame-level representations from the source and reference utterances.
\ours{} jointly defines shared states from these representations and computes source and reference statistics within the same states; no linguistic interpretation of the states is assumed.
For each state, \ours{} estimates a source-to-reference mean shift in the original WavLM space.
Source-frame posterior probabilities then combine these state-specific shifts to produce frame-level updates without training or fine-tuning neural parameters.

\subsection{Shared-State Definition and Assignment}
\label{sec:shared_partition}
The frozen WavLM encoder \(F\) extracts
$
H_q
=
F(x_q)
\in
\mathbb{R}^{T_q\times D},
q\in\{s,r\},
$
where \(T_q\) is the number of frames in the source~$s$ or reference~$r$ and where~\(D\) is the representation dimension.

For each conversion, we fit the GMM using a reduced set of WavLM feature dimensions.
We rank the source feature dimensions by their temporal standard deviation,
$
v_d
=
\operatorname{Std}\!\left(H_s[:,d]\right).
$
We retain the \(d_c\) source dimensions with the largest temporal standard deviations and denote their indices by \(\mathcal{I}_s\).
We then select these same feature dimensions from the reference representation.
This gives source and reference representations expressed in the same \(d_c\)-dimensional coordinate system:
$
Z_q
=
H_q[:,\mathcal{I}_s]
\in
\mathbb{R}^{T_q\times d_c},
q\in\{s,r\}.
$
The reduced representations \(Z_q\) are used only for state definition and frame-to-state routing.
All state statistics and conversion shifts are estimated and applied in the original \(D\)-dimensional WavLM space.

Fitting separate GMMs to the source and reference utterances would define two independent sets of components and would therefore require an additional component-matching step.
We instead concatenate the source and reference frames and fit one GMM to the resulting pooled representations.
Each mixture component therefore defines one shared state for both utterances.
We use a diagonal covariance parameterization:
\begin{equation}
p(z)=
\sum_{k=1}^{K}
\pi_k
\mathcal{N}
\left(
z;
\boldsymbol{\nu}_k,
\operatorname{diag}(\boldsymbol{\sigma}_k^2)
\right).
\end{equation}

As usual, for each state \(k\), \(\pi_k\) denotes the mixture weight, \(\boldsymbol{\nu}_k\) the component mean, and \(\boldsymbol{\sigma}_k^2\) the diagonal variance. 
We then fit candidate GMMs with number of components \(K\le K_{\max}\).
For model selection, each pooled frame \(Z_q[i]\) is temporarily assigned to
$
k_q^\star(i)
=
\arg\max_k
p(k\mid Z_q[i]).
$
We discard any GMM candidate for which a state contains fewer than \(n_{\min}\) assigned pooled frames.
Among the remaining candidates, the GMM with the lowest Bayesian information criterion (BIC)~\cite{schwartz1978estimating} is selected, and its number of states is denoted by \(K_{\mathrm e}\).

The selected GMM is then used to assign each frame a posterior probability over the set of
\(K_{\mathrm{e}}\) states:
\begin{equation}
\gamma^{(q)}_{i,k}
=
p(k\mid Z_q[i]),
\quad
q\in\{s,r\},
\quad
k=1,\ldots,K_{\mathrm{e}}.
\end{equation}
We refer to \(\gamma^{(q)}_{i,k}\) as the routing probability from frame \(i\) of utterance \(q\) to state \(k\). 

To reduce frame-to-frame variation in the routing probabilities, we apply temporal smoothing over a window of \(L=2J+1\) frames:
\begin{equation}
\bar{\gamma}_{i,k}^{(q)}
=
\frac{1}{L}
\sum_{j=-J}^{J}
\gamma_{c_q(i+j),k}^{(q)},
\qquad
\hat{\gamma}_{i,k}^{(q)}
=
\frac{
\bar{\gamma}_{i,k}^{(q)}
}{
\sum_{\ell=1}^{K_{\mathrm e}}
\bar{\gamma}_{i,\ell}^{(q)}
+
\varepsilon
},
\label{eq:smoothed_posteriors}
\end{equation}
where \(c_q(u)=\min\{\max\{u,1\},T_q\}\) clips out-of-range indices to the nearest valid frame and \(\varepsilon\) ensures numerical stability.

\subsection{Local Shift Estimation}
\label{sec:state_shifts}

For each state, \ours{} estimates a source-to-reference mean shift in the original \(D\)-dimensional WavLM space.
The state statistics are computed using the smoothed routing probabilities \(\hat{\gamma}_{i,k}^{(q)}\).

For state \(k\), the total posterior weight and full-dimensional weighted mean are
\begin{equation}
M_{q,k}
=
\sum_{i=1}^{T_q}
\hat{\gamma}_{i,k}^{(q)},
\quad
\boldsymbol{\mu}_{q,k}
=
\frac{
\sum_{i=1}^{T_q}
\hat{\gamma}_{i,k}^{(q)}H_q[i]
}{
M_{q,k}
},
\quad
q\in\{s,r\}.
\label{eq:state_statistics}
\end{equation}

For a state whose total posterior mass is at least \(m_{\min}\) in both utterances, we define the source-to-reference shift as the difference between the corresponding state means.
Otherwise, we use the utterance-level mean difference:
\begin{equation}
\boldsymbol{\Delta}_k^\star
=
\begin{cases}
\boldsymbol{\mu}_{r,k}
-
\boldsymbol{\mu}_{s,k},
&
M_{s,k}\geq m_{\min}
\ \text{and}\
M_{r,k}\geq m_{\min},
\\[1mm]
\boldsymbol{\mu}_{r}^{\mathrm g}
-
\boldsymbol{\mu}_{s}^{\mathrm g},
&
\text{otherwise},
\end{cases}
\label{eq:state_displacement}
\end{equation}
where
\[
\boldsymbol{\mu}_{q}^{\mathrm g}
=
\frac{1}{T_q}
\sum_{i=1}^{T_q}
H_q[i].
\]

The state-specific shift controls the source-to-reference change within each local region, while the utterance-level mean difference provides a fallback when a local estimate is poorly supported.

\subsection{Posterior-Weighted Frame Update}
\label{sec:frame_update}

Given the estimated local shifts, each source frame is updated by a posterior-weighted combination:
\begin{equation}
H'_s[t]
=
H_s[t]
+
\lambda
\sum_{k=1}^{K_{\mathrm e}}
\hat{\gamma}_{t,k}^{(s)}
\boldsymbol{\Delta}_k^\star,
\label{eq:statevc}
\end{equation}
where \(\lambda\in[0,1]\) controls the conversion strength.
We refer to this posterior-weighted combination of state-specific shifts as \emph{soft routing}.
Because the posterior weights depend on the source frame, different frames can receive different combinations of the state-specific shifts.

Finally, a frozen WavLM-conditioned HiFi-GAN~\cite{kong2020hifigan} vocoder released with kNN-VC~\cite{baas2023knnvc}\footnote{\url{https://github.com/bshall/knn-vc}}
reconstructs the converted waveform from \(H'_s\).

\begin{figure*}[t]
    \centering
    \includegraphics[width=0.95\textwidth]{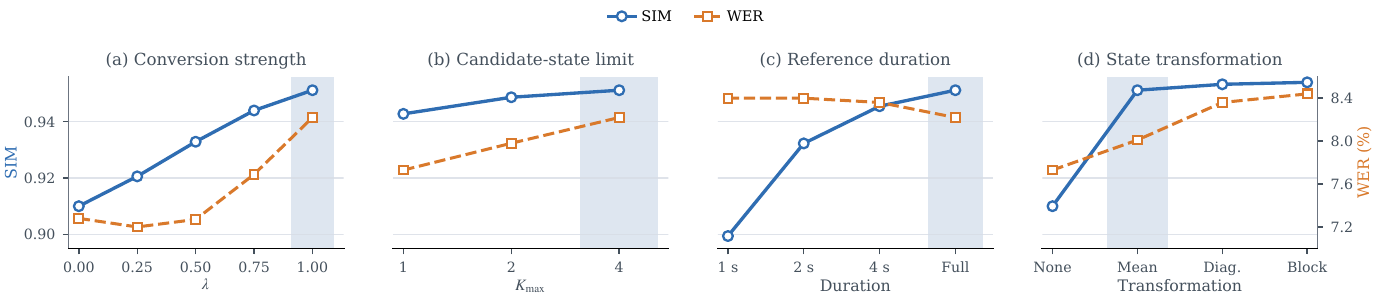}
    \caption{
    Sensitivity and design analyses for
    (a) conversion strength \(\lambda\),
    (b) candidate-state limit \(K_{\max}\),
    (c) reference duration, and
    (d) state transformation.
    SIM and WER use the left and right axes, respectively.
    Panels (a--c) use Mean, while panel (d) compares state-specific transformations.
    Shaded regions indicate the selected settings; comparisons are made within each panel.
    }
    \label{fig:sensitivity_design}
\end{figure*}

\section{Experimental Setup}
\label{sec:setup}

Following MKL-VC~\cite{lobashev2025mklvc}, we construct 7,800 cross-speaker pairs from LibriSpeech \texttt{test-clean}~\cite{panayotov2015librispeech}: five source utterances from each of 40 speakers, each paired with one reference utterance from every other speaker.
All systems use the same source--reference pairs.
We include FreeVC and FreeVC-S~\cite{li2023freevc} as learned zero-shot references and compare \ours{} primarily with the training-free methods kNN-VC~\cite{baas2023knnvc}, LinearVC~\cite{kamper2025linearvc}, MKL-VC~\cite{lobashev2025mklvc}, \(k\)DOT~\cite{selitskiy2026dotvc}, USCF~\cite{li2026uscf}, and SSL-GMMVC~\cite{tanabu2026sslgmmvc}.
For consistency with our evaluation setting, we adopt the configurations from the original papers that are compatible with our setup.
We use MKL-VC with \(k=2\), \(k\)DOT with \(k=4\), USCF with \(W_1\) and \(r=75\), and SSL-GMMVC CrossDiag with \(K=1\).

Speaker similarity is measured by cosine similarity (SIM) between SpeechBrain x-vectors\footnote{\url{https://huggingface.co/speechbrain/spkrec-xvect-voxceleb}} extracted from the converted and reference waveforms~\cite{snyder2018xvectors}.
Content preservation is evaluated for all systems using the same Whisper-base\footnote{\url{https://huggingface.co/openai/whisper-base}} WER/CER protocol against the source transcript~\cite{radford2023whisper}.
Twenty listeners each evaluate the outputs of all compared VC systems for the same 40 source--reference pairs.
Naturalness is rated on a 1--5 scale (N-MOS), and perceived similarity to the reference speaker on a 1--4 scale (S-MOS).
Reported scores are averaged over listeners and source--reference pairs for each system.

\ours{} uses frozen sixth-layer WavLM-Large representations with \(D=1024\), \(d_c=24\), \(K_{\max}=4\), \(n_{\min}=20\), \(m_{\min}=20\), \(\lambda=1\), and posterior smoothing with \(L=5\) frames (\(J=2\)).
Here, \(n_{\min}\) is the minimum assigned-frame count used to validate a candidate GMM, whereas \(m_{\min}\) is the minimum posterior mass required to use a state-specific shift.

To examine whether the shared-state framework requires transformations more flexible than a mean shift, we compare three state-specific transformations.
Mean applies only the state-specific shift in Eq.~\eqref{eq:state_displacement}.
Diagonal additionally rescales each source-centered coordinate by the corresponding reference-to-source state-wise standard-deviation ratio before recentering at the reference mean.
Block further incorporates within-block covariance through Gaussian Monge--Kantorovich linear transport~\cite{takatsu2011wasserstein} over contiguous 2-D feature blocks.
All three transformations act in the full \(D=1024\) WavLM space, while \(d_c=24\) is used only to define the shared states and compute routing probabilities.

For the separate-GMM state-definition comparison, source and reference states are matched by the Hungarian algorithm~\cite{kuhn1955hungarian}.
For the Block transformation, we compare hard and soft frame-level routing.
Both use the same smoothed state posteriors, state means and covariances, and state-specific Block transformations.
Hard routing selects the Block-transformed output of the most probable state, \(k^\star(t)=\arg\max_k\hat{\gamma}_{t,k}^{(s)}\), whereas soft routing combines the state-specific Block-transformed outputs using the source-frame posterior probabilities.

Code and audio samples are available at
\url{https://github.com/eurecom-asp/StateVC}
and
\url{https://github.com/eurecom-asp/StateVC_listening_samples}.
\begin{table}[t]
\caption{Objective and subjective results. WER/CER are percentages.}
\label{tab:main_results}
\centering
\scriptsize
\renewcommand{\arraystretch}{0.95}
\begin{tabular*}{0.92\columnwidth}{@{\extracolsep{\fill}}lccccc@{}}
\toprule
Method & SIM \(\uparrow\) & WER \(\downarrow\) & CER \(\downarrow\) & N-MOS \(\uparrow\) & S-MOS \(\uparrow\)\\
\midrule
\multicolumn{6}{@{}l}{\emph{Learned zero-shot}}\\
FreeVC   & 0.9473 & 8.82 & 3.70 & 3.800 & 2.600\\
FreeVC-S & 0.9441 & 8.63 & 3.60 & \textbf{3.850} & 2.500\\
\addlinespace[1pt]
\multicolumn{6}{@{}l}{\emph{Training-free}}\\
LinearVC   & \textbf{0.9753} & 47.65 & 28.69 & 2.500 & 2.200\\
\(k\)DOT   & 0.9600 & 39.52 & 23.41 & 2.150 & 1.600\\
kNN-VC     & 0.9699 & 37.50 & 22.17 & 2.850 & 2.200\\
USCF       & 0.9515 & 20.36 & 11.38 & 2.250 & 1.850\\
SSL-GMMVC  & 0.9303 & 11.87 & 5.39  & 1.850 & 1.650\\
MKL-VC     & 0.9456 & 8.22 & 3.37  & 3.100 & 2.450\\
\midrule
\textbf{\ours}
& 0.9512 & \textbf{8.01} & \textbf{3.22} & 3.550 & \textbf{2.750}\\
\bottomrule
\end{tabular*}
\end{table}

\begin{table}[t]
\caption{State-definition and routing comparisons with the Block transformation fixed.}
\label{tab:state_routing}
\centering
\scriptsize
\renewcommand{\arraystretch}{0.95}
\begin{tabular*}{0.92\columnwidth}{@{\extracolsep{\fill}}lccc@{}}
\toprule
Variant & SIM \(\uparrow\) & WER \(\downarrow\) & CER \(\downarrow\)\\
\midrule
\multicolumn{4}{@{}l}{\emph{State definition}}\\
Separate GMMs + same-index & 0.9467 & 40.28 & 26.40\\
Separate GMMs + Hungarian & \textbf{0.9541} & 13.82 & 6.60\\
Shared GMM & 0.9540 & \textbf{8.44} & \textbf{3.48}\\
\addlinespace[1pt]
\multicolumn{4}{@{}l}{\emph{Frame-level routing}}\\
Hard: single-state Block transform
& \textbf{0.9556} & 9.14 & 3.86\\

Soft: posterior-weighted Block transforms
& 0.9540 & \textbf{8.44} & \textbf{3.48}\\
\bottomrule
\end{tabular*}
\end{table}

\section{Results}
\label{sec:results}

Table~\ref{tab:main_results} reports the overall VC results.
On the 7,800 LibriSpeech cross-speaker pairs, \ours{} achieves 8.01\% WER and 3.22\% CER, the lowest values among the evaluated systems, with a SIM of 0.9512.
Compared with MKL-VC, \ours{} reduces WER from 8.22\% to 8.01\% and CER from 3.37\% to 3.22\%, while increasing SIM from 0.9456 to 0.9512.
It also achieves the highest mean S-MOS among the evaluated systems and the highest mean N-MOS among the evaluated training-free systems.

Table~\ref{tab:state_routing} evaluates state definition and frame-level routing with the Block transformation fixed.
Compared with Hungarian matching of separately fitted states, the shared-state definition reduces WER from 13.82\% to 8.44\% and CER from 6.60\% to 3.48\%, at nearly identical SIM (0.9541 vs.\ 0.9540).
With the shared states, state statistics, and Block transformations fixed, soft routing reduces WER from 9.14\% to 8.44\% and CER from 3.86\% to 3.48\%, while SIM decreases slightly from 0.9556 to 0.9540.

Figure~\ref{fig:sensitivity_design} reports the state-transformation and sensitivity analyses.
Figure~\ref{fig:sensitivity_design}(d) shows that more flexible state-specific transformations provide only small additional SIM gains beyond Mean, while WER increases from 8.01\% for Mean to 8.36\% for Diagonal and 8.44\% for Block.
Mean captures most of the observed SIM gain over None, with a smaller WER increase than the more flexible transformations.
We therefore use Mean in the final system.
Figure~\ref{fig:sensitivity_design}(a--c) further shows that SIM increases with \(\lambda\), increasing \(K_{\max}\) slightly improves SIM at the cost of higher WER, and longer reference speech mainly improves SIM.

\section{Conclusions}
\label{sec:conclusion}

We presented \ours{}, a training-free one-shot VC method that jointly defines shared local states for source and reference representations and applies state-specific mean shifts through posterior-weighted frame updates.
Under the LibriSpeech one-shot protocol, \ours{} achieves the lowest WER and CER among the evaluated systems, at 8.01\% and 3.22\%, respectively, with a SIM of 0.9512.
Controlled comparisons show that jointly defined states yield lower recognition errors than separately fitted and matched states at nearly identical SIM, while a simple mean shift captures most of the SIM improvement obtained with more flexible transformations.
The current evaluation is limited to English LibriSpeech under the reported one-shot protocol.

\section*{Acknowledgment}
This work was supported by the French Agence Nationale de la Recherche (ANR) through the SpeechPrivacy project (ANR-23-CE23-0022).
ChatGPT was used solely for language editing and stylistic refinement.
All scientific claims, experimental results, analyses, and conclusions were independently reviewed and verified by the authors.

\bibliographystyle{IEEEbib}
\bibliography{refs}

\end{document}